\documentclass[amsmath,amssymb,twocolumn,aps,prb,superscriptaddress,floatfix,10pt]{revtex4-2}

\usepackage{graphicx}
\usepackage{ifthen}
\usepackage{xcolor}
\usepackage{bm}
\usepackage{braket}
\usepackage{multirow}
\usepackage{hyperref}
\usepackage{soul}
\usepackage[normalem]{ulem}
\usepackage{multirow}
\hypersetup{
 pdfnewwindow=true, colorlinks=true,
 linkcolor=blue, anchorcolor=blue,
 citecolor=blue, filecolor=blue,
 menucolor=blue, urlcolor=blue}

\graphicspath{{./}{figures/}}

\def\ve{\varepsilon}
\def\tc{T_{\rm c}}
\def\efeff{E_{\rm F}^{\rm eff}}
\def\ef{\ve_{\rm F}}
\def\bk{\mathbf{k}}
\def\bq{\mathbf{q}}
\def\o{\omega}
\def\op{\omega^\prime}

\def\l{\lambda}
\def\lV{\lambda^{\rm V}}
\def\D{\Delta}
\def\NF{N_{\rm F}}
\def\wV{w_{\rm V}}
\newcommand{\wlg}{\ensuremath{\omega_{\log}}}
\newcommand{\aaff}{\ensuremath{\alpha^2F(\omega)}}
\newcommand{\aaffv}{\ensuremath{\alpha^2F^{\rm V}(\o,\op)}}

\begin{document}

\title{Superconductivity in Noncentrosymmetric NbReSi: Beyond Harmonic and Adiabatic Limits}

\author{Shashi B. Mishra}
\email{mshashi125@gmail.com}
\affiliation{Laboratory for Physical Sciences, University of Maryland, College Park, MD 20740, USA}
\author{Sohair ElMeligy}
\affiliation{Department of Physics, Virginia Tech, Blacksburg, Virginia 24061, USA}
\author{Pratibha Dev}
\email{pdev@lps.umd.edu}
\affiliation{Laboratory for Physical Sciences, University of Maryland, College Park, MD 20740, USA}
\date{\today}

\begin{abstract}
The noncentrosymmetric superconductor NbReSi ($\tc=6.5$~K) is being actively explored for unconventional pairing that is allowed by its broken inversion symmetry. Despite broad experimental interest, what drives its superconductivity remains an open question. We show that the electronic states at the Fermi level are dominated by Nb and Re $d$-states.  Vibrations of the same heavy-metal framework contribute more than $95\%$ of the electron-phonon coupling, with the strongest coupling arising from the sublattice-selective breathing mode involving strongly-bonded Re-atoms. Harmonic Migdal-Eliashberg theory overestimates $\tc$ by more than $60\%$. The zero-point anharmonic hardening of these modes reduces the coupling by $20\%$. The leading vertex correction becomes important as well despite low phonon energies due to the contribution of spatially localized Re-$d$ orbitals to the electronic band structure near the Fermi level. Together, anharmonicity and vertex corrections bring the calculated $\tc$ and superconducting gap into close agreement with experiment, establishing NbReSi as a moderately coupled, phonon-mediated superconductor whose quantitative description requires going beyond harmonic Migdal--Eliashberg theory.
\end{abstract}

\maketitle


\section{\label{sec:intro}Introduction}

First identified as a superconductor in the mid-1980s~\cite{Subba1985}, there is an increased interest in the noncentrosymmetric compound NbReSi~\cite{Su2021,Shang2022,Sajilesh2022,Nandi2023,Chakrabortty2025} as its broken inversion symmetry allows for unconventional pairing~\cite{Bauer2012,Smidman2017}. In such materials, antisymmetric spin-orbit coupling (ASOC) lifts the spin degeneracy of the bands near the Fermi level~\cite{Gorkov2001,Frigeri2004}, allowing spin-singlet and -triplet pairings to mix. A sufficiently strong triplet component relative to the singlet component (usually quantified by $E_{\text{ASOC}}/k_{\text{B}}\tc$) can give rise to unconventional gap structures, time-reversal-symmetry breaking, and/or topological superconductivity~\cite{Smidman2017,Sato2017}. NbReSi provides a platform for examining these effects without the strong electronic correlations of $3d$ compounds~\cite{Subba1985,Shang2022,Singh2023}.

Experiments report two polymorphs of NbReSi: the hexagonal ZrNiAl-type structure ($P\bar{6}2m$, No.~189)~\cite{Su2021,Shang2022,Nandi2023} and the orthorhombic FeSiTi-type structure ($Ima2$, No.~46)~\cite{Subba1985,Sajilesh2022,Chakrabortty2025}. Both exhibit bulk type-II superconductivity with transition temperatures near $6.5$~K and upper critical fields $\mu_0H_{c2}(0)\approx11.5$--$13.5$~T, close to the Pauli limit~\cite{Su2021,Sajilesh2022,Shang2022,Nandi2023,Chakrabortty2025}. For the hexagonal phase considered here, polycrystalline samples consistently give $\tc\simeq6.5$~K~\cite{Su2021,Shang2022}, while a subsequent single-crystal study reported $\tc=6.1$~K~\cite{Nandi2023}. Muon-spin-rotation and NMR measurements indicate a fully gapped state that preserves time-reversal symmetry, although weak gap anisotropy or multigap behavior have not been completely excluded~\cite{Shang2022,Chakrabortty2025}.  For the magnetic field applied within the $ab$ plane and along the $c$-axis of a single-crystal, the measurements further reveal only modest upper-critical-field anisotropy, $H_{c2}^{ab}/H_{c2}^{c}\approx1.4$~\cite{Nandi2023}, indicating that the superconductivity is at least quasi-3D and mostly isotropic~\cite{Sajilesh2022,Shang2022}. 
Together, these observations suggest that ASOC does not generate a substantial triplet component in this Re-compound. NbReSi, therefore, provides a useful test case for determining whether its superconducting properties can be quantitatively described within an electron-phonon framework and for assessing the extent to which spin-orbit coupling modifies that description. 

On the theoretical side, there are no investigations to determine the microscopic origin of superconductivity in NbReSi, although its dynamical and electronic structure properties were studied in an earlier work~\cite{Basak2023}. In particular, it remains unclear whether the observed $\tc$ can be explained by conventional electron--phonon coupling and which phonon and electronic states dominate the pairing.

In this paper, we present a first-principles study of superconductivity in hexagonal NbReSi. We find that NbReSi is a conventional phonon-mediated superconductor, with more than 95\% of the electron--phonon coupling originating from low-energy Nb- and Re-dominated vibrations. An accurate quantitative description, however, requires going beyond the harmonic Migdal-Eliashberg theory. Anharmonic renormalization strongly modifies the vibrations that carry the largest coupling and involve strongly-bonded Re-atoms, while the lowest-order electron--phonon vertex correction~\cite{Mishra2025} remains non-negligible even for a metal with low phonon energies, which we attribute to contributions from spatially localized Re-$d$ orbitals to the band structure at the Fermi level. Together, these corrections bring the calculated $\tc$ and superconducting gap into agreement with experiment, whereas spin--orbit coupling changes the coupling strength only by $4\%$.

\begin{figure*}[!t]
    \centering
    \includegraphics[width=\linewidth]{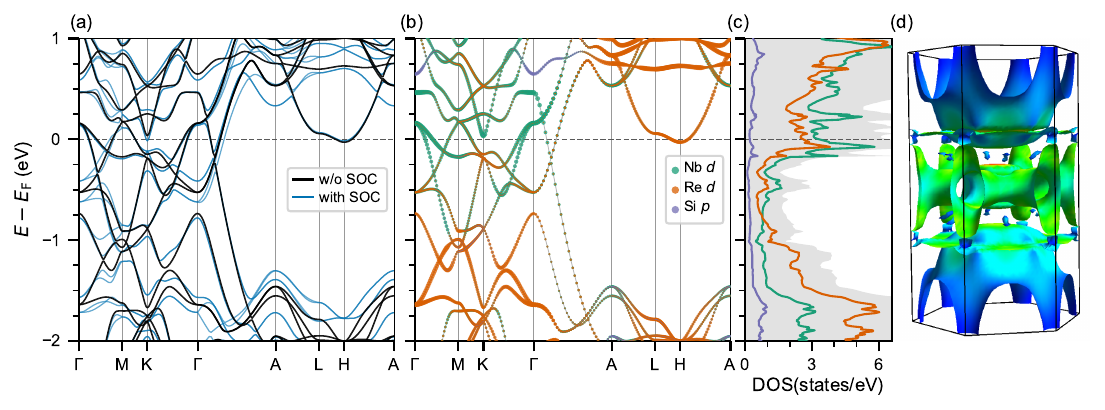}
    \caption{Electronic structure of NbReSi. (a) Band structure without (black) and with (blue) SOC; the dashed line marks the Fermi level $\ef$. (b) Orbital-projected bands; marker size is proportional to the Nb-$d$ (teal), Re-$d$ (orange), and Si-$p$ (purple) weight. (c) Total and orbital-resolved density of states per unit cell. (d) Fermi surface without SOC, colored by the Fermi velocity (blue: low, green: high).}
    \label{fig:bands}
\end{figure*}

\section{\label{sec:methods}Computational Methods}

Density-functional theory (DFT) calculations were performed with Quantum {\small ESPRESSO}~\cite{Giannozzi2017,Baroni2001}, using optimized norm-conserving Vanderbilt pseudopotentials~\cite{Hamann2013} from the Pseudo Dojo library~\cite{Vansetten2018} and the Perdew-Burke-Ernzerhof exchange-correlation functional~\cite{Perdew1996}. A plane-wave cutoff of $80$~Ry, Methfessel-Paxton smearing of $0.01$~Ry, and a $\Gamma$-centered $6\times6\times12$ $\bk$-grid were used. Lattice parameters and internal coordinates were relaxed to $10^{-4}$~Ry in energy and $10^{-3}$~Ry/\AA{} in force. Dynamical matrices and the linear variation of the self-consistent potential were computed within density-functional perturbation theory (DFPT) on a $4\times4\times8$ $\bq$-mesh, using both scalar-relativistic and fully relativistic pseudopotentials to assess the role of spin-orbit coupling.

Electron-phonon matrix elements were interpolated in the Wannier representation with the EPW code~\cite{Giustino2007,Ponce2016,Lee2023}. The Wannier functions were constructed from Nb-$d$, Re-$d$, and Si-$p$ projections ($39$ maximally localized functions) on a $\Gamma$-centered $4\times4\times8$ $\bk$-grid~\cite{Marzari2012,Pizzi2020}, and the interpolation was carried out onto $40\times40\times80$ $\bk$ and $20\times20\times40$ $\bq$ grids, retaining states within $\pm0.2$~eV of the Fermi level. Electronic and phononic Dirac deltas were replaced by Gaussians of $50$ and $0.5$~meV width. The accuracy of the interpolation is verified in Fig.~S1, and the convergence of $\l$, \wlg{}, and the gap with the fine grids in Fig.~S2 and Table~S1~\cite{SI}.

The isotropic Eliashberg equations were solved in the Fermi-surface-restricted (FSR) and full-bandwidth (FBW) formulations~\cite{Margine2013,Lee2023,Lucrezi2024,Mishra2025} on the imaginary axis with a Matsubara cutoff of $0.6$~eV. Nonadiabatic effects enter through the lowest-order electron-phonon vertex correction implemented in EPW~\cite{Mishra2025}, which introduces the two-frequency spectral function \aaffv{} and the vertex coupling $\lV=4\!\int\!\!\int\aaffv/(\o\op)\,d\o\,d\op$. 
We used a Coulomb pseudopotential ($\mu^*$) value of $0.21$, which has previously been used for elemental Nb~\cite{Savrasov1996}.

Anharmonic phonons were computed with the special displacement method (A-SDM) implemented in the ZG module of EPW~\cite{Zacharias2023,Lee2023}, using a finite displacement of $0.1$~\AA{} in a $2\times2\times2$ supercell, iterated to self-consistency.

\section{\label{sec:results}Results and Discussion}

\subsection{\label{sec:elec}Crystal and electronic structure}

Hexagonal structure of NbReSi ($P\bar{6}2m$, No.~189) has nine atoms per cell: Nb at $3f$ $(x_{\rm Nb},0,0)$ with $x_{\rm Nb}\approx 0.60$, Re at $3g$ $(x_{\rm Re},0,\tfrac12)$ with $x_{\rm Re}\approx 0.26$, and Si at $2d$ $(\tfrac13,\tfrac23,\tfrac12)$ and $1a$ $(0,0,0)$. Structural relaxation yields lattice parameters of $a=6.87$~\AA{} and $c=3.32$~\AA{}, in close agreement with the experimental values ($a=6.72$~\AA{} and $c=3.48$~\AA{})~\cite{Su2021}.  The two metal sublattices are geometrically distinct in the $ab$ plane [Fig.~S3~\cite{SI}]. The Nb atoms form a distorted kagome-like lattice at $z=0$, with corner-sharing equilateral triangles of side $3.62$~\AA{} and a Si atom at the center of the distorted hexagons. On the other hand, Re atoms are arranged in corner-sharing equilateral triangles of two different sizes, forming a highly distorted kagome-like net at $z=\tfrac12$. The sides of the two equilateral triangles are given by the Re--Re distances of $4.48$~\AA{} and $3.07$~\AA{}, with the latter forming strongly bonded Re trimers (hereafter referred to as the Re$_3$ units), while the Re atoms in the larger triangles are not bonded.  The Si atoms in the $z=\tfrac12$ plane act as scaffolding atoms and sit in the centers of the highly distorted hexagons and the larger equilateral triangles, forming Re--Si bonds of length $2.58$~\AA, which stabilize the Re's kagome motif. The Re$_3$ units are threaded by the Si $1a$ chain along $c$ (Re--Si $=2.43$~\AA, the shortest bond).

Figure~\ref{fig:bands}(a) compares the electronic bands without and with SOC. The largest SOC splittings, about $0.2$~eV at $\Gamma$- and A-points, agree with the earlier report~\cite{Su2021}. These splittings arise in the bands that are mostly derived from the $d$-orbitals of the heavy elements -- Nb and Re.  The Kramers' degeneracy is preserved at $\Gamma$ and A since these are time-reversal-invariant-momenta points. The bands crossing $\ef$ split by less than $25$~meV, so ASOC leaves the Fermi-level electronic structure essentially unchanged and modifies the electron-phonon coupling by only $4\%$ (Fig.~S4~\cite{SI}), indicating its minor role in superconducting pairing, if any.
The projected bands and density of states (DOS) [Fig.~\ref{fig:bands}(b,c)] show that the states at $\ef$ are dominated by Nb- and Re-$d$ orbitals, with a small Si-$p$ admixture. Nb-$d$ character prevails in the $k_z=0$ plane along $\Gamma$--M--K--$\Gamma$, and mostly Re-$d$ character in the $k_z=\pi/c$ plane along A--L--H--A, reflecting the crystalline sublattices at $z=0$ and $z=\tfrac12$. A Wannier tight-binding analysis (Appendix~\ref{app:hop}) shows that Nb--Re hybridization across the two layers creates these states, while strong in-plane Re--Re hopping within the trimers pushes the band below $\ef$. The Fermi surface [Fig.~\ref{fig:bands}(d)] consists of several three-dimensional sheets, with $N(\ef)=2.33$ states/eV per formula unit (both spins), in good agreement with the experimental estimate of $2.10$ obtained from the Sommerfeld coefficient after removing the electron-phonon enhancement, $N(\ef)=3\gamma_n/[\pi^2k_{\rm B}^2(1+\l)]$, using $\gamma_n=8.23$~mJ,mol$^{-1}$K$^{-2}$ and $\l=0.66$~\cite{Su2021} (Table~\ref{tab:supercond}).

\begin{figure}[!t]
    \centering
    \includegraphics[width=\linewidth]{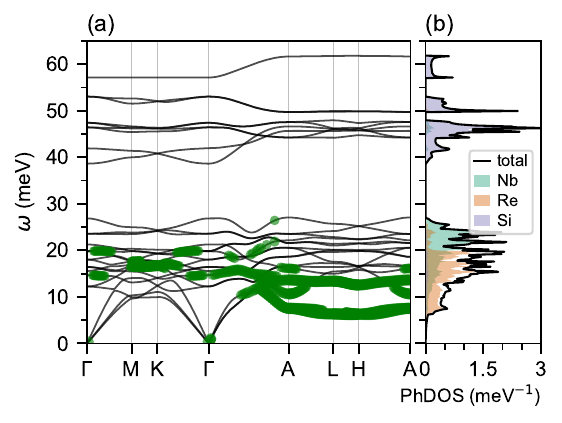}
    \caption{Harmonic phonons in NbReSi. (a) Phonon dispersion with circle size proportional to the mode-resolved coupling $\lambda_{\mathbf{q}\nu}$. (b) Total and atom-projected phonon DOS.}
    \label{fig:phonons}
\end{figure}
\subsection{\label{sec:phonons}Phonons and electron-phonon coupling}

\begin{figure}[!t]
    \centering
    \includegraphics[width=\linewidth]{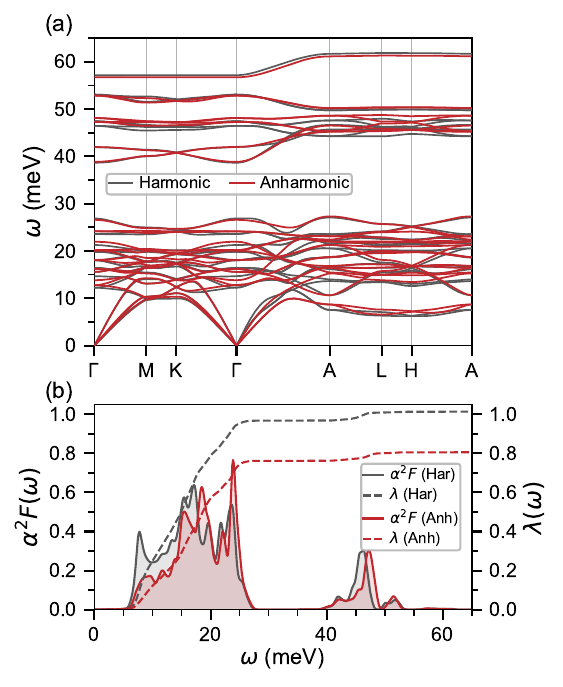}
    \caption{Anharmonic renormalization of the lattice dynamics. (a) Phonon dispersion with harmonic (black) and anharmonic (red) approximations. (b) Eliashberg spectral function \aaff{} (solid, left axis) and cumulative electron-phonon coupling $\l(\o)$ (dashed, right axis).}
    \label{fig:anh}
\end{figure}

All phonon modes are real throughout the Brillouin zone [Fig.~\ref{fig:phonons}], confirming the dynamical stability of the hexagonal phase~\cite{Basak2023}. The mass contrast opens a $\sim12\text{ meV}$ gap, isolating the high-frequency Si optical modes ($>39$~meV) from the lower-lying Nb/Re manifold ($<27$~meV), which transitions from mostly Re-based vibrations to Nb dominance with increasing energy. The mode-resolved coupling $\l_{\bq\nu}$ is concentrated in the three lowest Re-dominated branches, particularly along A--L--H--A, where the acoustic modes soften near the zone boundary. The modes below $27$~meV supply more than $95\%$ of $\l$ [see $\l(\o)$ in Fig.~\ref{fig:anh}(b)], while Si modes contribute negligibly. The electron-phonon coupling is thus governed by vibrations of the same Nb--Re framework that provides the states at $\ef$.

\subsection{\label{sec:anh}Anharmonic renormalization}

Because the coupling is concentrated in soft low-energy modes, we next assess anharmonic effects arising from zero-point and thermal lattice fluctuations~\cite{Zacharias2023,Mishra2025,Belli2025}. Anharmonicity produces only modest changes in the phonon spectrum [Fig.~\ref{fig:anh}(a,b)], primarily hardening the low-energy Nb--Re branches while leaving the Si modes nearly unchanged. This hardening, nevertheless, redistributes \aaff{} away from the lowest energies, reducing $\l$ from $1.01$ to $0.80$ and increasing \wlg{} from $168$ to $190$~K [Fig.~\ref{fig:anh}(b)]. This renormalization arises primarily from zero-point motion, with thermal effects producing only minor changes (Figs.~S5 and S6~\cite{SI}).

\begin{figure}[!t]
    \centering
    \includegraphics[width=0.95\linewidth]{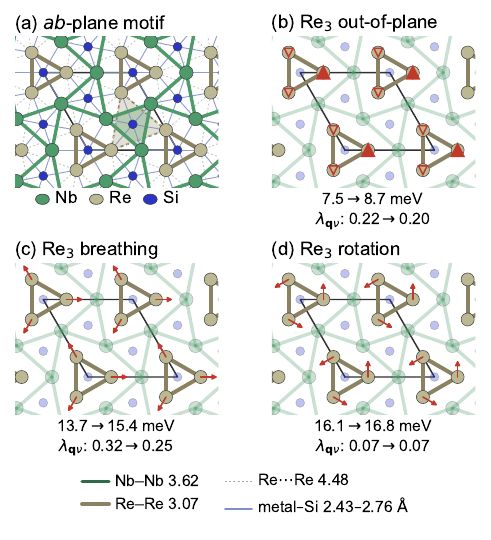}
    \caption{Microscopic origin of the anharmonic correction. (a) Nb and Re sublattices in the $ab$ plane, showing distorted Nb kagome net and isolated Re$_3$ trimers; dashed lines mark Re$\cdots$Re contacts between triangles. (b)--(d) Representative Re-dominated modes at A $=(0,0,\tfrac12)$, corresponding to (b) out-of-plane motion, (c) in-plane breathing, and (d) in-plane rotation. Filled/open triangles in (b) denote opposite out-of-plane displacements. Labels give the harmonic$\to$anharmonic frequencies and $\l_{\bq\nu}$.}
    \label{fig:re3}
\end{figure}

The microscopic origin of this renormalization lies in the Re$_3$ trimer motif (Fig.~\ref{fig:re3}). Its nearly dispersionless breathing branch at $12.6$--$13.8$~meV in the $k_z=\pi/c$ plane carries the strongest coupling, $\l_{\bq\nu}=0.29$--$0.33$ at A, L, and H. Anharmonicity hardens this branch by $1.6$--$2.1$~meV and reduces $\l_{\bq\nu}$ to $0.22$--$0.25$ (Table~S2~\cite{SI}); the lower-energy out-of-plane Re$_3$ vibrations harden by up to $1.2$~meV with a weaker, wave-vector-dependent change of $\l_{\bq\nu}$. By contrast, the weakly coupled Re$_3$ rotation and the Si modes remain essentially unchanged. Anharmonicity, therefore, selectively renormalizes the Re$_3$ vibrations that dominate the pairing, explaining why modest phonon shifts reduce the total $\l$ by $20\%$. The weak in-plane Re--Re force constants further identify the trimer breathing coordinate as particularly soft, with its restoring force governed primarily by Re--Si and Re--Nb interactions (Table~S3~\cite{SI}).

\begin{figure}[!t]
    \centering
    \includegraphics[width=\linewidth]{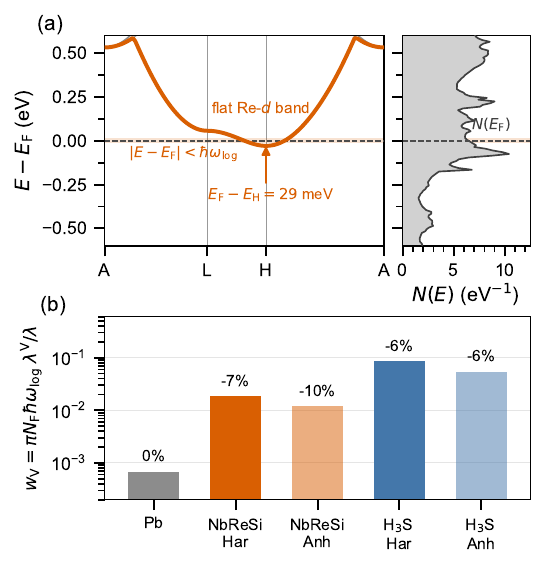}
    \caption{Electronic origin of the nonadiabatic correction. (a) Bands along A--L--H--A and corresponding DOS. The nearly dispersionless Re-$d$ band reaches its minimum $29$~meV below $\ef$ at H; shading marks $|E-\ef|<\hbar\wlg$. (b) Nonadiabatic weight $\wV$ for Pb, NbReSi, and H$_3$S (Appendix~\ref{app:wv}; Pb and H$_3$S from Ref.~\cite{Mishra2025}). Labels give the vertex-induced relative change $\delta\tc/\tc=(\tc^{\rm FBW+ver}-\tc^{\rm FBW})/\tc^{\rm FBW}$.}
    \label{fig:vertex_origin}
\end{figure}

\subsection{\label{sec:vertex}Electron-phonon vertex correction}

Beyond anharmonic lattice dynamics, we examine the lowest-order vertex correction~\cite{Mishra2025} [Fig.~S7~\cite{SI}], which captures the nonadiabatic processes neglected in Migdal--Eliashberg theory~\cite{Allen1983}. Migdal's theorem~\cite{Migdal1958} justifies their neglect when $\l\hbar\o_0/\ef\ll1$, and with $\hbar\wlg\approx15$~meV would ordinarily suggest that such corrections are negligible in NbReSi. 
Here, however, the relevant electronic scale is set by a nearly dispersionless Re-$d$ band along L--H (in the $k_z=\pi/c$ plane), whose minimum at H lies only $29$~meV below $\ef$ [Fig.~\ref{fig:vertex_origin}(a)]. This band produces the sharp variation of $N(E)$ near $\ef$, and $N(E)$ changes by a factor of three within $0.3$~eV, which sets an effective electronic scale $\efeff=0.1$--$0.3$~eV. The Fermi-surface-averaged velocity is correspondingly small, $\langle v_{\rm F}\rangle=2.1\times10^{5}$~m/s, comparable to the $(2$--$4)\times10^{5}$~m/s inferred for H$_3$S~\cite{Talantsev2022,Mishra2026,Mishra_hydrides}
and an order of magnitude below the free-electron value of Pb~\cite{Lykken1970}, and provides an extended phase space of states within $\hbar\wlg$ of the Fermi level (Appendix~\ref{app:wv}).

In the isotropic theory, the vertex term of the Eliashberg equations carries an explicit factor of $\NF$~\cite{Mishra2025}, so that its weight relative to the Migdal term is measured by $\wV=\pi\NF\hbar\wlg\lV/\l$, i.e.\ by the vertex ratio $\lV/\l$ of Ref.~\cite{Mishra2026,Mishra_hydrides} weighted by $\pi\NF\hbar\wlg$. 
Although the vertex ratio is modest, $\lV/\l=0.12$ (harmonic) and $0.07$ (anharmonic), the large $\NF$ yields $\wV=0.02$, intermediate between Pb ($7\times10^{-4}$), where vertex effects are negligible~\cite{Mishra2025}, and H$_3$S ($0.05$--$0.09$) [Fig.~\ref{fig:vertex_origin}(b); see also Appendix~\ref{app:wv}]. Together with $\eta=\hbar\wlg/\efeff\approx0.1$, this places NbReSi in the weakly nonadiabatic regime~\cite{Mishra_hydrides}, where vertex corrections refine rather than control $\tc$. Nevertheless, a $2$--$3\%$ change in the pairing kernel lowers $\tc$ by $7$--$10\%$, amplified by the strong sensitivity of $\tc$ to the coupling at moderate $\l$ ($d\ln\tc/d\l\approx5$ for $\l=0.8$ and $\mu^*=0.21$). Anharmonic hardening of the strongly coupled Re$_3$ modes further weakens the vertex contribution, reducing $\lV$ from $0.12$ to $0.05$ (Fig.~S8~\cite{SI}).

\begin{figure}[!t]
    \centering
    \includegraphics[width=\linewidth]{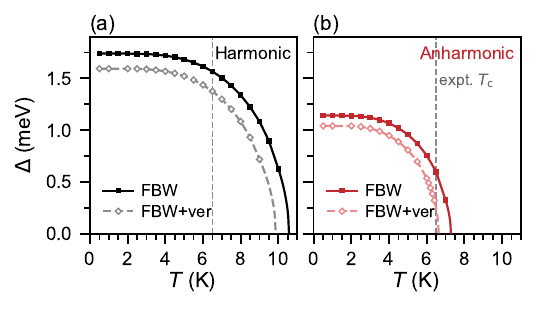}
    \caption{Isotropic superconducting gap $\D(T)$ of NbReSi at $\mu^*=0.21$ for (a) harmonic and (b) anharmonic phonons, at the FBW (solid) and FBW+ver (dashed) levels. The vertical line marks the experimental $\tc=6.5$~K of hexagonal phase~\cite{Su2021,Shang2022}.}
    \label{fig:sc}
\end{figure}

\begingroup
\squeezetable
\begin{table}[!t]
\centering
\caption{Superconducting properties of NbReSi for harmonic (Har) and anharmonic (Anh) phonons, compared with experiment (Expt). $N(\ef)$ is the bare DOS per formula unit (both spins) and \wlg{} is in K. $\tc$ values are isotropic Eliashberg results at $\mu^*=0.21$ in the FSR and FBW formulations, without and with the lowest-order vertex correction ($+$ver); $\D(0)$ and $2\D(0)/k_{\rm B}\tc$ refer to FBW${+\rm ver}$. Experimental $N(\ef)$, $\tc$, $\D(0)$, and $\l$ are for hexagonal polycrystals, from Refs.~\cite{Su2021,Shang2022}.}
\label{tab:supercond}
\setlength{\tabcolsep}{0.7pt}
\begin{ruledtabular}
\begin{tabular}{l c ccc cccc cc}
 & $N(\ef)$ &\multicolumn{3}{c}{e-ph coupling} & \multicolumn{4}{c}{$\tc$ (K)} & $\D(0)$ & $\frac{2\D(0)}{k_{\rm B}\tc}$ \\
\cline{3-5}\cline{6-9}
 & (eV$^{-1}$) &$\l$ & $\lV$ & \wlg{} & FSR & FBW & FSR${+\rm ver}$ & FBW${+\rm ver}$ &  &  \\
\colrule
Har & 2.33 & 1.01 & 0.12 & 168 & 10.9 & 10.6 & 9.9 & 9.9 & 1.59 & 3.75 \\
Anh & 2.32 & 0.80 & 0.05 & 190 & 7.8 & 7.3 & 6.8 & 6.6 & 1.04 & 3.65 \\
Expt & 2.1 & 0.66 & -- & -- & \multicolumn{4}{c}{6.5} & 1.0 & 3.5--3.9 \\
\end{tabular}
\end{ruledtabular}
\end{table}
\endgroup

\subsection{\label{sec:sc}Superconductivity}

We solve the isotropic Eliashberg equations with harmonic and anharmonic phonons, with and without the vertex correction, in the FSR and FBW formulations (Fig.~\ref{fig:sc}, Table~\ref{tab:supercond}). At the harmonic Migdal level, the FBW calculation gives $\tc=10.6$~K, well above the experimental $\tc\simeq6.5$~K of hexagonal NbReSi~\cite{Su2021,Shang2022}. Anharmonicity provides the dominant correction, reducing $\tc$ to $7.3$~K, while the vertex correction lowers it further to $6.6$~K, in close agreement with experiment. Full-bandwidth effects are small, with FSR and FBW differing by less than $0.5$~K. The anharmonic vertex-corrected gap, $\D(0)=1.04$~meV, and ratio $2\D(0)/k_{\rm B}\tc=3.65$ are consistent with $\mu$SR and NMR measurements~\cite{Shang2022,Sajilesh2022}. Although the absolute $\tc$ depends on $\mu^*$, the gap ratio and the hierarchy of corrections are robust over $\mu^*=0.10$--$0.21$ (Fig.~S9 and Table~S4~\cite{SI}). The weak gap anisotropy and Nb/Re site mixing reported experimentally~\cite{Nandi2023,Chakrabortty2025} lie beyond the present isotropic treatment and warrant a future anisotropic Eliashberg study~\cite{Margine2013,Mishra2024}.

\section{\label{sec:conclusions}Conclusions}

NbReSi is a moderately coupled, fully gapped phonon-mediated superconductor in which the Nb--Re framework supplies both the Fermi-level states and the pairing phonons, while spin-orbit coupling plays a minor role. Harmonic Migdal--Eliashberg theory overestimates $\tc$ by more than $60\%$. Two corrections with a common microscopic origin restore agreement with experiment: zero-point anharmonic hardening of the Re$_3$ trimer vibrations that carry the strongest coupling, and the electron-phonon vertex correction, which is non-negligible for a metal with $15$~meV characteristic phonons because a nearly dispersionless Re-$d$ band at the Fermi level produces a large $\NF$ and a small Fermi velocity. The nonadiabatic weight $\wV=\pi\NF\hbar\wlg\lV/\l$ complements the vertex and adiabaticity ratios of Ref.~\cite{Mishra2026} as a criterion for when vertex corrections matter in low-phonon-energy metals, and the same Re$_3$ physics should govern the isostructural ZrNiAl-type superconductors.

\begin{acknowledgments}
The authors acknowledge the Texas Advanced Computing Center (TACC) at The University of Texas at Austin (http://www.tacc.utexas.edu) for providing computational resources that have contributed to the research results reported within this paper.
\end{acknowledgments}

\appendix

\section{\label{app:hop}Electronic origin of the Fermi-level states}

\begin{figure}[!t]
    \centering
    \includegraphics[width=0.95\linewidth]{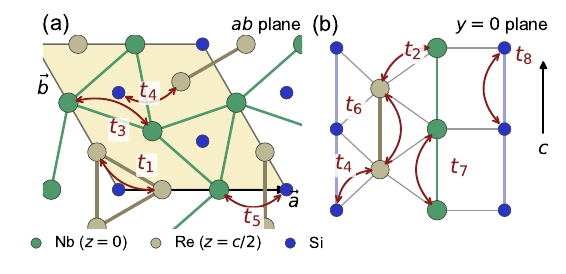}
    \caption{Symmetry-distinct hopping channels of Table~\ref{tab:hop}. (a) $ab$ plane showing the Nb kagome-like net (green, $z=0$), Re$_3$ trimers ($z=c/2$), and Si; the shaded rhombus marks the unit cell. The in-plane channels are $t_1$ (Re--Re), $t_3$ (Nb--Nb), $t_4$ [Re--Si$(2d)$], and $t_5$ [Nb--Si$(1a)$]. (b) $y=0$ plane showing the $c$-axis channels $t_6$ (Re--Re), $t_7$ (Nb--Nb), and $t_8$ (Si--Si), and the interlayer channels $t_2$ (Nb--Re) and $t_4$ [Re--Si$(1a)$].}
    \label{fig:hop}
\end{figure}

Figure~\ref{fig:hop} marks the symmetry-distinct hopping channels $t_1$--$t_8$ of the 39-orbital Wannier Hamiltonian, while Table~\ref{tab:hop} lists their largest matrix elements and the resulting change in $N(\ef)$ when one channel is removed at fixed band filling. Suppressing the interlayer Nb--Re hopping $t_2$ more than doubles $N(\ef)$, while removing the Nb--Nb chain hopping $t_7$ increases it by about a quarter, showing that these channels disperse the metal-$d$ manifold and control the position of the flat Re-$d$ band in Fig.~\ref{fig:vertex_origin}(a). In contrast, removing the intra-trimer hopping $t_1$ reduces $N(\ef)$ by $64\%$, because strong bonding within the Re$_3$ trimers places their antibonding Re-$d$ states near $\ef$. The large Fermi-level weight therefore emerges from Re$_3$ trimers hybridized with the Nb kagome net, directly linking the electronic states at $\ef$ to the structural units whose breathing modes dominate the pairing.

\begingroup
\squeezetable
\begin{table}[!t]
\centering
\caption{Hopping channels of Fig.~\ref{fig:hop}: bond length $d$, largest matrix element $|t|_{\max}$, and change $\delta N(\ef)$ when the channel is removed at fixed band filling ($32\times32\times64$ $\bk$-mesh, $30$~meV Gaussian broadening). A negative $\delta N(\ef)$ marks a channel that creates states at $\ef$.}
\label{tab:hop}
\setlength{\tabcolsep}{2pt}
\begin{ruledtabular}
\begin{tabular}{l l c c r}
 & Bond & $d$ (\AA) & $|t|_{\max}$ (eV) & $\delta N(\ef)$ \\
\colrule
$t_1$ & Re--Re trimer          & 3.07 & 0.75 & $-64\%$ \\
$t_2$ & Nb--Re                 & 2.86/2.94 & 0.86 & $+130\%$ \\
$t_3$ & Nb--Nb kagome          & 3.62 & 0.50 & $+7\%$ \\
$t_4$ & Re--Si                 & 2.43/2.58 & 1.36 & $-34\%$ \\
$t_5$ & Nb--Si                 & 2.67/2.76 & 1.17 & $-19\%$ \\
$t_6$ & Re--Re $\parallel c$   & 3.32 & 0.86 & $-33\%$ \\
$t_7$ & Nb--Nb $\parallel c$   & 3.32 & 0.84 & $+27\%$ \\
$t_8$ & Si--Si $\parallel c$   & 3.32 & 1.51 & $-3\%$ \\
\end{tabular}
\end{ruledtabular}
\end{table}
\endgroup

\begin{figure}[!t]
    \centering
    \includegraphics[width=0.95\linewidth]{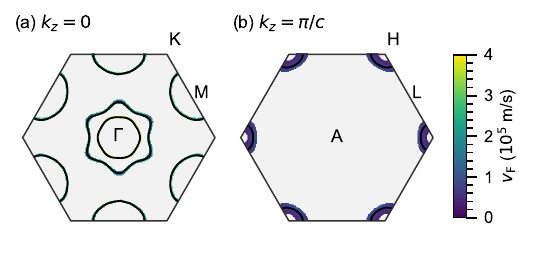}
    \caption{Fermi-surface cuts (black) in the (a) $k_z=0$ and (b) $k_z=\pi/c$ planes. The colored region is the shell of states within $\hbar\wlg=14.5$~meV of $\ef$, into which a phonon can scatter an electron on shell, colored by the Fermi velocity; it is broad and slow around H, where the Re-$d$ band is nearly dispersionless.}
    \label{fig:fsplanes}
\end{figure}

\section{\label{app:wv}Nonadiabatic weight}

\begingroup
\squeezetable
\begin{table}[!t]
\centering
\caption{Nonadiabatic weight $\wV=\pi\NF\hbar\wlg\lV/\l$ and the vertex-induced change of the FBW $\tc$, $\delta\tc/\tc$. $\NF$ is per spin and unit cell; $\l$, \wlg{}, and $\lV$ are the converged values of Table~\ref{tab:supercond}. Pb and H$_3$S results are from Ref.~\cite{Mishra2025}.}
\label{tab:nonadia}
\setlength{\tabcolsep}{2pt}
\begin{ruledtabular}
\begin{tabular}{l c c c c c c c c}
 & phonon & $\NF$ & $\hbar\wlg$ & $\l$ & $\lV$ & $\lV/\l$ & $\wV$ & $\delta\tc/\tc$ \\
&  & (eV$^{-1}$) & (meV) & & & & & \\
\colrule
Pb   & Har           & 0.27 & 5.4  & 1.29 & 0.20 & 0.15 & 0.0007 & $0$ \\
NbReSi & Har      & 3.50 & 14.5 & 1.01 & 0.12 & 0.12 & 0.018 & $-7\%$ \\
NbReSi & Anh      & 3.48 & 16.4 & 0.80 & 0.05 & 0.07 & 0.012 & $-10\%$ \\
H$_3$S & Har      & 0.33 & 107  & 2.29 & 1.78 & 0.78 & 0.085 & $-6\%$ \\
H$_3$S & Anh      & 0.33 & 127  & 1.72 & 0.71 & 0.41 & 0.054 & $-6\%$ \\
\end{tabular}
\end{ruledtabular}
\end{table}
\endgroup

The vertex spectral weight \aaffv{} lies below $30$~meV, within the Nb/Re phonon manifold [Fig.~S8~\cite{SI}]. In the Fermi-surface-restricted formulation, the vertex contribution to the gap equation is~\cite{Mishra2025}

\begin{align}\label{eq:fsr-vertex}
\D(i\o_j)Z(i\o_j)&=\pi k_{\rm B}T\sum_{j'}\big[\l(i\o_j-i\o_{j'})-\mu^*\big]\tilde\gamma^{\D}(i\o_{j'})\nonumber\\
&+\pi^{3}(k_{\rm B}T)^{2}\NF\sum_{j'j''}\lV_{jj'j''}\,\tilde{\gamma}^{T}_{j'}\tilde{P}^{\D}\tilde{\gamma}_{j''},
\end{align}
with an analogous term in the equation for $Z$ and notation as in Ref.~\cite{Mishra2025}. Since each Matsubara sum is restricted to the phonon scale, the vertex-to-Migdal ratio is characterized, up to factors of order unity, by
\begin{equation}
\wV=\pi\NF\hbar\wlg\,\frac{\lV}{\l}.
\end{equation}
The product $\NF\lV$, and hence $\wV$, is independent of unit-cell choice, allowing comparison between materials; the quantities entering $\wV$ are listed in Table~\ref{tab:nonadia}.

The enhanced $\wV$ in NbReSi originates from the low-velocity states near H [Fig.~\ref{fig:fsplanes}], where the nearly flat Re-$d$ band produces a broad electronic shell within $\hbar\wlg$ of $\ef$. Consistently, the effective electronic scale is only $\efeff\simeq0.1$--$0.3$~eV, giving $\eta=\hbar\wlg/\efeff\simeq0.05$--$0.13$. The normalized four-state on-shell phase space, $C(0)/\NF^4$, increases from $1.0$ over the full Fermi surface to $4.4$ near H [Fig.~S7 and Sec.~S5~\cite{SI}], identifying the enhancement with the flat-band region rather than global Fermi-surface nesting.

\bibliography{pap}

@misc{SI,
note = {{See Supplemental Material for computational details and k/q-grid convergence (S2), convergence and temperature dependence of the anharmonic phonons (S3), the effect of spin-orbit coupling (S4), the dependence on $\mu^*$ (S5), and the Fermi velocities and on-shell phase space underlying the nonadiabatic correction (S6).}}
}

@article{Savrasov1996,
  title = {Electron-phonon interactions and related physical properties of metals from linear-response theory},
  author = {Savrasov, S. Y. and Savrasov, D. Y.},
  journal = {Phys. Rev. B},
  volume = {54},
  issue = {23},
  pages = {16487--16501},
  numpages = {0},
  year = {1996},
  month = {Dec},
  publisher = {American Physical Society},
  doi = {10.1103/PhysRevB.54.16487},
  url = {https://link.aps.org/doi/10.1103/PhysRevB.54.16487}
}

@article{Lykken1970,
  title = {{Measurement of the Fermi Velocity in Single-Crystal Films of Lead by Electron Tunneling}},
  author = {Lykken, G. I. and Geiger, A. L. and Mitchell, E. N.},
  journal = {Phys. Rev. Lett.},
  volume = {25},
  issue = {22},
  pages = {1578--1580},
  numpages = {0},
  year = {1970},
  month = {Nov},
  publisher = {American Physical Society},
  doi = {10.1103/PhysRevLett.25.1578},
  url = {https://link.aps.org/doi/10.1103/PhysRevLett.25.1578}
}

@article{Mishra_hydrides,
      title={{Anharmonicity and Nonadiabaticity in Hydride Superconductors}}, 
      author={Shashi B. Mishra and Francesco Belli and Eva Zurek and Elena R. Margine},
      year={2026},
      journal={arXiv:2608.14949},
      url={https://arxiv.org/abs/2608.14949}
}

@article{Mishra2026,
author = {Mishra, Shashi B. and Margine, Elena R.},
title = {{Nonadiabatic and Anharmonic Effects in High-Pressure H$_3$S and D$_3$S Superconductors}},
journal = {Annalen der Physik},
volume = {538},
number = {1},
pages = {e00553},
doi = {https://doi.org/10.1002/andp.202500553},
url = {https://onlinelibrary.wiley.com/doi/abs/10.1002/andp.202500553},
year = {2026}
}

@article{Migdal1958,
  title={Interaction between electrons and lattice vibrations in a normal metal},
  author={Migdal, A. B.},
  journal={Sov. Phys. JETP},
  volume={7},
  pages={996-1001},
  year={1958},
  url={http://jetp.ras.ru/cgi-bin/dn/e_007_06_0996.pdf}
}

@article{Mishra2024,
  title = {{Stability-superconductivity map for compressed Na-intercalated graphite}},
  author = {Mishra, Shashi B. and Marcial, Edan T. and Debata, Suryakanti and Kolmogorov, Aleksey N. and Margine, Elena R.},
  journal = {Phys. Rev. B},
  volume = {110},
  issue = {17},
  pages = {174508},
  numpages = {13},
  year = {2024},
  month = {Nov},
  publisher = {American Physical Society},
  doi = {10.1103/PhysRevB.110.174508},
  url = {https://link.aps.org/doi/10.1103/PhysRevB.110.174508}
}

@article{Lee2023,
  title = {Electron–phonon physics from first principles using the {EPW} code},
  author = {Lee, Hyungjun and Poncé, Samuel and Bushick, Kyle and  Hajinazar, Samad and Lafuente-Bartolome, Jon and Leveillee, Joshua and Lian, Chao and Lihm, Jae-Mo and Macheda, Francesco and Mori, Hitoshi and Paudyal, Hari and Sio, Weng Hong and Tiwari, Sabyasachi and Zacharias, Marios and Zhang, Xiao and Bonini, Nicola and Kioupakis, Emmanouil and Margine, Elena R. and Giustino, Feliciano},
  journal = {npj Comput. Mater.},
  volume = {9},
  issue = {1},
  pages = {2057-3960},
  year = {2023},
  url = {https://doi.org/10.1038/s41524-023-01107-3},
  doi = {10.1038/s41524-023-01107-3}
}

@incollection{Allen1983,
title = {{Theory of Superconducting $T_{\rm c}$}},
editor = {Henry Ehrenreich and Frederick Seitz and David Turnbull},
series = {Solid State Physics},
publisher = {Academic Press},
volume = {37},
pages = {1-92},
year = {1983},
url = {https://www.sciencedirect.com/science/article/pii/S0081194708606657},
author = {Philip B. Allen and Božidar Mitrović}
}

@article{Allen1975,
  title = {Transition temperature of strong-coupled superconductors reanalyzed},
  author = {Allen, P. B. and Dynes, R. C.},
  journal = {Phys. Rev. B},
  volume = {12},
  issue = {3},
  pages = {905--922},
  numpages = {0},
  year = {1975},
  month = {Aug},
  publisher = {American Physical Society},
  doi = {10.1103/PhysRevB.12.905},
  url = {https://link.aps.org/doi/10.1103/PhysRevB.12.905}
}

@article{Lucrezi2024,
  title={{Full-bandwidth anisotropic Migdal-Eliashberg theory and its application to superhydrides}},
  author={Lucrezi, Roman and Ferreira, Pedro P and Hajinazar, Samad and Mori, Hitoshi and Paudyal, Hari and Margine, Elena R and Heil, Christoph},
  journal={Commun. Phys.},
  volume={7},
  number={1},
  pages={33},
  year={2024},
  publisher={Nature Publishing Group UK London},
  doi = {10.1038/s42005-024-01528-6},
  url = {https://doi.org/10.1038/s42005-024-01528-6}
}

@article{Hamann2013,
  title = {{Optimized norm-conserving Vanderbilt pseudopotentials}},
  author = {Hamann, D. R.},
  journal = {Phys. Rev. B},
  volume = {88},
  issue = {8},
  pages = {085117},
  numpages = {10},
  year = {2013},
  month = {Aug},
  publisher = {American Physical Society},
  doi = {10.1103/PhysRevB.88.085117},
  url = {https://link.aps.org/doi/10.1103/PhysRevB.88.085117}
}

@article{Giannozzi2017,
  author = {Giannozzi, Paolo and Andreussi, Oliviero and Brumme, Thomas and Bunau, Olivier and Buongiorno Nardelli, Marco and Calandra, Matteo and Car, Roberto and Cavazzoni, Carlo and Ceresoli, Davide and Cococcioni, Matteo and Colonna, Nicola and Carnimeo, Ivan and Dal Corso, Andrea and de Gironcoli, Stefano and Delugas, Pietro and DiStasio Jr, Robert A. and Ferretti, Andrea and Floris, Andrea and Fratesi, Guido and Fugallo, Giorgio and Gebauer, Ralph and Gerstmann, Uwe and Giustino, Feliciano and Gorni, Tommaso and Jia, Jia and Kawamura, Masayuki and Ko, Hyo Yoon and Kokalj, Anton and Küçükbenli, Esen and Lazzeri, Michele and Marsili, Matteo and Marzari, Nicola and Mauri, Francesco and Nguyen, Ngoc Linh and Nguyen, Huy-Viet and Otero-de-la-Roza, Alberto and Paulatto, Lorenzo and Poncé, Samuel and Rocca, Dario and Sabatini, Riccardo and Santra, Biswajit and Schlipf, Martin and Seitsonen, Ari P. and Smogunov, Alexander and Timrov, Iurii and Thonhauser, Timo and Umari, Paolo and Vast, Nathalie and Wu, Xifan and Baroni, Stefano},
journal = {J. Phys.: Condens. Matter},
doi = {10.1088/1361-648x/aa8f79},
number = {46},
pages = {465901},
title = {{Advanced capabilities for materials modelling with Quantum ESPRESSO}},
volume = {29},
year = {2017},
  url = {https://doi.org/10.1088/1361-648x/aa8f79}
}

@article{Vansetten2018,
title = {{The PseudoDojo: Training and grading a 85 element optimized norm-conserving pseudopotential table}},
journal = {Comput. Phys. Commun.},
volume = {226},
pages = {39-54},
year = {2018},
issn = {0010-4655},
doi = {https://doi.org/10.1016/j.cpc.2018.01.012},
url = {https://www.sciencedirect.com/science/article/pii/S0010465518300250},
author = {M.J. {van Setten} and M. Giantomassi and E. Bousquet and M.J. Verstraete and D.R. Hamann and X. Gonze and G.-M. Rignanese}
}

@article{Perdew1996,
  title = {Generalized Gradient Approximation Made Simple},
  author = {Perdew, John P. and Burke, Kieron and Ernzerhof, Matthias},
  journal = {Phys. Rev. Lett.},
  volume = {77},
  issue = {18},
  pages = {3865--3868},
  numpages = {0},
  year = {1996},
  month = {Oct},
  publisher = {American Physical Society},
  doi = {10.1103/PhysRevLett.77.3865},
  url = {https://link.aps.org/doi/10.1103/PhysRevLett.77.3865}
}

@article{Marzari2012,
  title = {{Maximally localized Wannier functions: Theory and applications}},
  author = {Marzari, Nicola and Mostofi, Arash A. and Yates, Jonathan R. and Souza, Ivo and Vanderbilt, David},
  journal = {Rev. Mod. Phys.},
  volume = {84},
  issue = {4},
  pages = {1419--1475},
  numpages = {0},
  year = {2012},
  month = {Oct},
  publisher = {American Physical Society},
  doi = {10.1103/RevModPhys.84.1419},
  url = {https://link.aps.org/doi/10.1103/RevModPhys.84.1419}
}

@article{Pizzi2020,
	doi = {10.1088/1361-648x/ab51ff},
	url = {https://doi.org/10.1088%2F1361-648x%2Fab51ff},
	year = 2020,
	month = {jan},
	publisher = {{IOP} Publishing},
	volume = {32},
	number = {16},
	pages = {165902},
        author = {Pizzi, Giovanni and Vitale, Valerio and Arita, Ryotaro and Bl{\"u}gel, Stefan and Freimuth, Frank and G{\'e}ranton, Guillaume and Gibertini, Marco and Gresch, Dominik and Johnson, Charles and Koretsune, Takashi and Iba{\~n}ez-Azpiroz, Julen and Lee, Hyungjun and Lihm, Jae-Mo and Marchand, Daniel and Marrazzo, Antimo and Mokrousov, Yuriy and Mustafa, Jamal I. and Nohara, Yoshiro and Nomura, Yusuke and Paulatto, Lorenzo and Ponc{\'e}, Samuel and Ponweiser, Thomas and Qiao, Junfeng and Th{\"o}le, Florian and Tsirkin, Stepan S. and Wierzbowska, Ma{\l}gorzata and Marzari, Nicola and Vanderbilt, David and Souza, Ivo and Mostofi, Arash A. and Yates, Jonathan R.},
	title = {Wannier90 as a community code: new features and applications},
	journal = {J. Phys. Condens. Matter}
}

@article{Baroni2001,
  title = {Phonons and related crystal properties from density-functional perturbation theory},
  author = {Baroni, Stefano and de Gironcoli, Stefano and Dal Corso, Andrea and Giannozzi, Paolo},
  journal = {Rev. Mod. Phys.},
  volume = {73},
  issue = {2},
  pages = {515--562},
  numpages = {0},
  year = {2001},
  month = {Jul},
  publisher = {American Physical Society},
  doi = {10.1103/RevModPhys.73.515},
  url = {https://link.aps.org/doi/10.1103/RevModPhys.73.515}
}

@article{Giustino2007,
  title = {{Electron-phonon interaction using Wannier functions}},
  author = {Giustino, Feliciano and Cohen, Marvin L. and Louie, Steven G.},
  journal = {Phys. Rev. B},
  volume = {76},
  issue = {16},
  pages = {165108},
  numpages = {19},
  year = {2007},
  month = {Oct},
  publisher = {American Physical Society},
  doi = {10.1103/PhysRevB.76.165108},
  url = {https://link.aps.org/doi/10.1103/PhysRevB.76.165108}
}

@article{Ponce2016,
title = {{{EPW}: Electron–phonon coupling, transport and superconducting properties using maximally localized Wannier functions}},
journal = {Comput. Phys. Commun.},
volume = {209},
pages = {116-133},
year = {2016},
issn = {0010-4655},
doi = {https://doi.org/10.1016/j.cpc.2016.07.028},
url = {https://www.sciencedirect.com/science/article/pii/S0010465516302260},
author = {S. Poncé and E.R. Margine and C. Verdi and F. Giustino}
}

@article{Margine2013,
  title = {{Anisotropic Migdal-Eliashberg theory using Wannier functions}},
  author = {Margine, E. R. and Giustino, F.},
  journal = {Phys. Rev. B},
  volume = {87},
  issue = {2},
  pages = {024505},
  numpages = {12},
  year = {2013},
  month = {Jan},
  publisher = {American Physical Society},
  doi = {10.1103/PhysRevB.87.024505},
  url = {https://link.aps.org/doi/10.1103/PhysRevB.87.024505}
}

@article{Zacharias2023,
  title = {Anharmonic lattice dynamics via the special displacement method},
  author = {Zacharias, Marios and Volonakis, George and Giustino, Feliciano and Even, Jacky},
  journal = {Phys. Rev. B},
  volume = {108},
  issue = {3},
  pages = {035155},
  numpages = {16},
  year = {2023},
  month = {Jul},
  publisher = {American Physical Society},
  doi = {10.1103/PhysRevB.108.035155},
  url = {https://link.aps.org/doi/10.1103/PhysRevB.108.035155}
}

@article{Talantsev2022,
    author = {Talantsev, Evgeny F.},
    title = {{Universal Fermi velocity in highly compressed hydride superconductors}},
    journal = {Matter Radiat. Extrem.},
    volume = {7},
    number = {5},
    pages = {058403},
    year = {2022},
    month = {08},
    issn = {2468-2047},
    doi = {10.1063/5.0091446},
    url = {https://doi.org/10.1063/5.0091446}
}

@article{Mishra2025,
  author={Shashi B. Mishra and Hitoshi Mori and Elena R. Margine},
  title = {{Electron-phonon vertex correction effect in superconducting H$_3$S}}, 
  journal = {npj Computational Materials},
  volume = {11},
  issue = {1},
  pages = {342},
  year = {2025},
  doi={10.1038/s41524-025-01818-9},
  url={https://doi.org/10.1038/s41524-025-01818-9}
}

@article{Belli2025,
  title={{Efficient modelling of anharmonicity and quantum effects in PdCuH$_2$ with machine learning potentials}},
  author={Belli, Francesco and Zurek, Eva},
  journal={npj Comput. Mater.},
  volume={11},
  number={1},
  pages={87},
  year={2025},
  publisher={Nature Publishing Group UK London},
  url={https://www.nature.com/articles/s41524-025-01553-1}
}

@article{Shang2022,
  title = {{Evidence of fully gapped superconductivity in NbReSi: A combined $\mu$SR and NMR study}},
  author = {Shang, T. and Tay, D. and Su, H. and Yuan, H. Q. and Shiroka, T.},
  journal = {Phys. Rev. B},
  volume = {105},
  pages = {144506},
  year = {2022},
  doi = {10.1103/PhysRevB.105.144506},
  url = {https://doi.org/10.1103/physrevb.105.144506}
}

@article{Sajilesh2022,
  title = {{Superconductivity in noncentrosymmetric NbReSi investigated by muon spin rotation and relaxation}},
  author = {{Sajilesh K. P.} and Motla, K. and Meena, P. K. and Kataria, A. and Patra, C. and {Somesh K.} and Hillier, A. D. and Singh, R. P.},
  journal = {Phys. Rev. B},
  volume = {105},
  pages = {094523},
  year = {2022},
  doi = {10.1103/PhysRevB.105.094523},
  url = {https://doi.org/10.1103/physrevb.105.094523}
}

@article{Basak2023,
   title={{Theoretical Study of Dynamical and Electronic Properties of Noncentrosymmetric Superconductor NbReSi}},
   volume={16},
   url={http://dx.doi.org/10.3390/ma16010078},
   number={1},
   journal={Materials},
   publisher={MDPI AG},
   author={Basak, Surajit and Ptok, Andrzej},
   year={2022},
   month=Dec, pages={78},
  doi = {10.3390/ma16010078}
}

@article{Singh2023,
  title = {{Fully gapped superconductivity and topological aspects of the noncentrosymmetric superconductor TaReSi}},
  author = {Singh, D. and Barker, J. A. T. and Thamizhavel, A. and Hillier, A. D. and Singh, R. P.},
  journal = {Phys. Rev. B},
  volume = {107},
  pages = {224504},
  year = {2023},
  doi = {10.1103/PhysRevB.107.224504},
  url = {https://doi.org/10.1103/physrevb.107.224504}
}

@article{Smidman2017,
  title = {{Superconductivity and spin-orbit coupling in non-centrosymmetric materials: a review}},
  author = {Smidman, M. and Sherif, M. B. and Yuan, H. Q. and Agterberg, D. F. and Sigrist, M.},
  journal = {Rep. Prog. Phys.},
  volume = {80},
  pages = {036501},
  year = {2017},
  doi = {10.1088/1361-6633/80/3/036501},
  url = {https://doi.org/10.1088/1361-6633/80/3/036501}
}

@book{Bauer2012,
  title = {{Non-Centrosymmetric Superconductors: Introduction and Overview}},
  author = {Bauer, Ernst and Sigrist, Manfred},
  year = {2012},
  publisher = {Springer},
  series = {Lecture Notes in Physics},
  volume = {847},
  doi = {10.1007/978-3-642-24624-1},
  url = {https://doi.org/10.1007/978-3-642-24624-1}
}

@article{Gorkov2001,
  title = {{Superconducting $2D$ system with lifted spin degeneracy: mixed singlet-triplet state}},
  author = {Gor'kov, L. P. and Rashba, E. I.},
  journal = {Phys. Rev. Lett.},
  volume = {87},
  pages = {037004},
  year = {2001},
  doi = {10.1103/PhysRevLett.87.037004},
  url = {https://doi.org/10.1103/physrevlett.87.037004}
}

@article{Frigeri2004,
  title = {{Superconductivity without inversion symmetry: MnSi versus CePt$_3$Si}},
  author = {Frigeri, P. A. and Agterberg, D. F. and Koga, A. and Sigrist, M.},
  journal = {Phys. Rev. Lett.},
  volume = {92},
  pages = {097001},
  year = {2004},
  doi = {10.1103/PhysRevLett.92.097001},
  url = {https://doi.org/10.1103/physrevlett.92.097001}
}

@article{Su2021,
  title = {{NbReSi: A noncentrosymetric superconductor with large upper critical field}},
  author = {Su, H. and Shang, T. and Du, F. and Chen, C. F. and Ye, H. Q. and Lu, X. and Cao, C. and Smidman, M. and Yuan, H. Q.},
  journal = {Phys. Rev. Mater.},
  volume = {5},
  issue = {11},
  pages = {114802},
  numpages = {8},
  year = {2021},
  month = {Nov},
  publisher = {American Physical Society},
  doi = {10.1103/PhysRevMaterials.5.114802},
  url = {https://link.aps.org/doi/10.1103/PhysRevMaterials.5.114802}
}

@article{Subba1985,
  title={{Structure and superconductivity studies on ternary equiatomic silicides, MM'Si}},
  author={Subba Rao, GV and Wagner, K and Balakrishnan, Geetha and Janaki, J and Paulus, W and Sch{\"o}llhorn, R and Subramanian, VS and Poppe, U},
  journal={Bulletin of Materials Science},
  volume={7},
  number={3},
  pages={215--228},
  year={1985},
  publisher={Springer},
  url={https://www.ias.ac.in/article/fulltext/boms/007/03-04/0215-0228},
  doi = {10.1007/BF02747575}
}

@article{Sato2017,
  title = {{Topological superconductors: a review}},
  author = {Sato, Masatoshi and Ando, Yoichi},
  journal = {Rep. Prog. Phys.},
  volume = {80},
  pages = {076501},
  year = {2017},
  doi = {10.1088/1361-6633/aa6ac7},
  url = {https://doi.org/10.1088/1361-6633/aa6ac7}
}

@article{Nandi2023,
  title = {Anisotropic properties of a noncentrosymmetric NbReSi superconducting single crystal},
  author = {Nandi, Suman and Sasmal, Souvik and Maity, Bishal Baran and Sharma, Vikash and Dwari, Gourav and Kulkarni, Ruta and Thamizhavel, A.},
  journal = {Phys. Rev. B},
  volume = {107},
  issue = {13},
  pages = {134518},
  year = {2023},
  doi = {10.1103/PhysRevB.107.134518},
  url = {https://link.aps.org/doi/10.1103/PhysRevB.107.134518}
}

@article{Chakrabortty2025,
  title = {Influence of nonmagnetic impurities on the upper critical field, critical current density, and vortex pinning mechanism of the noncentrosymmetric superconductor NbReSi},
  author = {Chakrabortty, S. and Mohapatra, N.},
  journal = {Phys. Rev. B},
  volume = {111},
  issue = {9},
  pages = {094511},
  numpages = {20},
  year = {2025},
  month = {Mar},
  publisher = {American Physical Society},
  doi = {10.1103/PhysRevB.111.094511},
  url = {https://link.aps.org/doi/10.1103/PhysRevB.111.094511}
}

\end{document}


\title{Supplemental Material: \\ Superconductivity in Noncentrosymmetric NbReSi: Beyond Harmonic and Adiabatic Limits}

\author{Shashi B. Mishra}
\email{mshashi125@gmail.com}
\affiliation{Laboratory for Physical Sciences, University of Maryland, College Park, MD 20740, USA}
\author{Sohair ElMeligy}
\affiliation{Department of Physics, Virginia Tech, Blacksburg, Virginia 24061, USA}
\author{Pratibha Dev}
\email{pdev@lps.umd.edu}
\affiliation{Laboratory for Physical Sciences, University of Maryland, College Park, MD 20740, USA}
\date{\today}

\maketitle

\section{\label{sec:grid_conv}Wannier interpolation and $\bk$/$\bq$-grid convergence}

\begin{figure}[!hbt]
    \centering
    \includegraphics[width=0.85\linewidth]{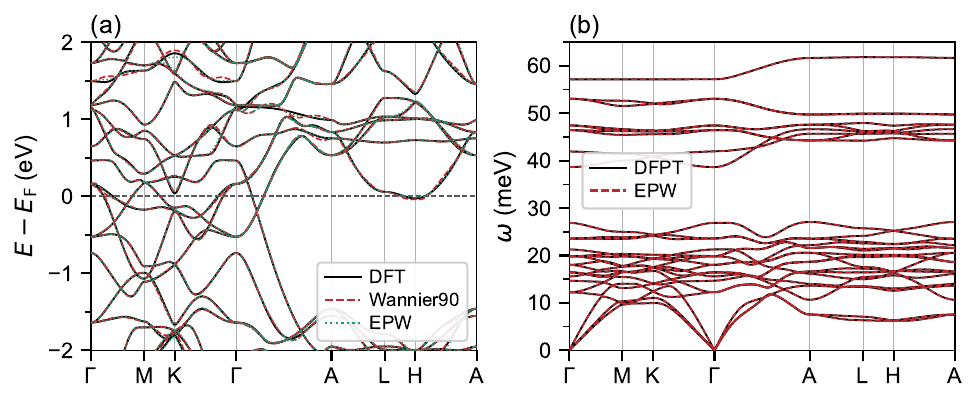}
    \caption{Wannier interpolation check. (a) DFT bands (black) and the bands interpolated from the 39 maximally localized Wannier functions by Wannier90 (red dashed) and EPW (green dotted). (b) DFPT phonons from the $4\times4\times8$ $\bq$-mesh (black) and the EPW Fourier interpolation (red dashed).}
    \label{fig:wannier_check}
\end{figure}

Figure~\ref{fig:wannier_check} compares the DFT bands and the DFPT phonons with their Wannier--Fourier interpolations, confirming the accuracy of the 39-orbital Wannier Hamiltonian used throughout.

\begin{figure}[!hbt]
    \centering
    \includegraphics[width=0.85\textwidth]{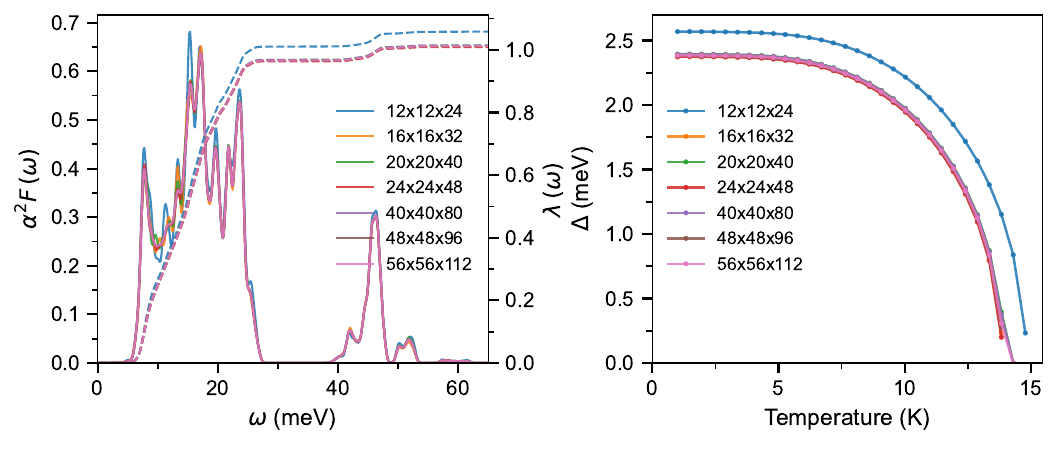}
    \caption{Convergence of the Eliashberg spectral function $\a^2F(\o)$ (solid, left axis) and cumulative coupling $\l(\o)$ (dashed, right axis) (left panel), and of the isotropic Fermi-surface-restricted gap $\Delta(T)$ at $\mu^*=0.10$ (right panel), with respect to the fine $\bk$/$\bq$ grids used in the EPW calculations, for harmonic phonons. Grids are labelled by the $\bk$-mesh; the $\bq$-mesh is halved along each direction.}
    \label{fig:grid_conv}
\end{figure}

Figure~\ref{fig:grid_conv} and Table~\ref{tab:grid} show the convergence of $\a^2F(\o)$, $\l$, $\o_{\log}$, and the isotropic gap with the fine $\bk$/$\bq$ grids; $\l$ is converged to within $0.5\%$ for all meshes of $16\times16\times32$ and denser.

\begin{table}[!hbt]
\centering
\caption{Convergence of the electron-phonon coupling $\l$ and the logarithmic average phonon frequency $\o_{\log}$ with the fine interpolation meshes for harmonic phonons. }
\label{tab:grid}
\setlength{\tabcolsep}{8pt}
\renewcommand{\arraystretch}{1.15}
\begin{tabular}{c c c c}
\hline\hline
$\bk$-mesh & $\bq$-mesh & $\l$ & $\o_{\log}$ (K) \\
\hline
$12\times12\times24$   & $6\times6\times12$   & 1.058 & 167.6 \\
$16\times16\times32$   & $8\times8\times16$   & 1.011 & 167.9 \\
$20\times20\times40$   & $10\times10\times20$ & 1.014 & 168.1 \\
$24\times24\times48$   & $12\times12\times24$ & 1.009 & 168.2 \\
$40\times40\times80$   & $20\times20\times40$ & 1.014 & 168.1 \\
$48\times48\times96$   & $24\times24\times48$ & 1.011 & 168.1 \\
$56\times56\times112$  & $28\times28\times56$ & 1.012 & 168.1 \\
\hline\hline
\end{tabular}
\end{table}

\section{\label{sec:structure}Crystal structure and effect of spin-orbit coupling}


\begin{figure}[!hbt]
    \centering
    \includegraphics[width=0.95\linewidth]{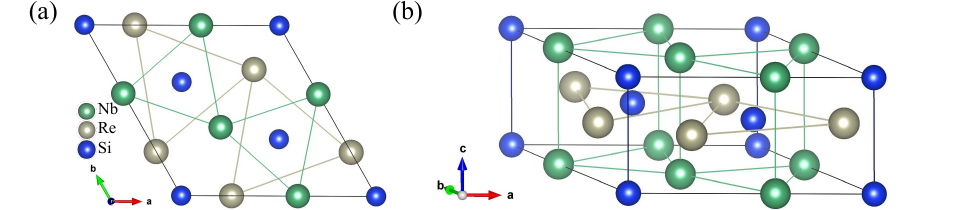}
    \caption{Crystal structure of NbReSi in the hexagonal ZrNiAl-type structure (space group $P\bar{6}2m$, No.~189): (a) top view and (b) side view. Nb, Re, and Si atoms are green, tan, and blue spheres, respectively.}
    \label{fig:crystal}
\end{figure}

\begin{figure}[!hbt]
    \centering
    \includegraphics[width=0.85\textwidth]{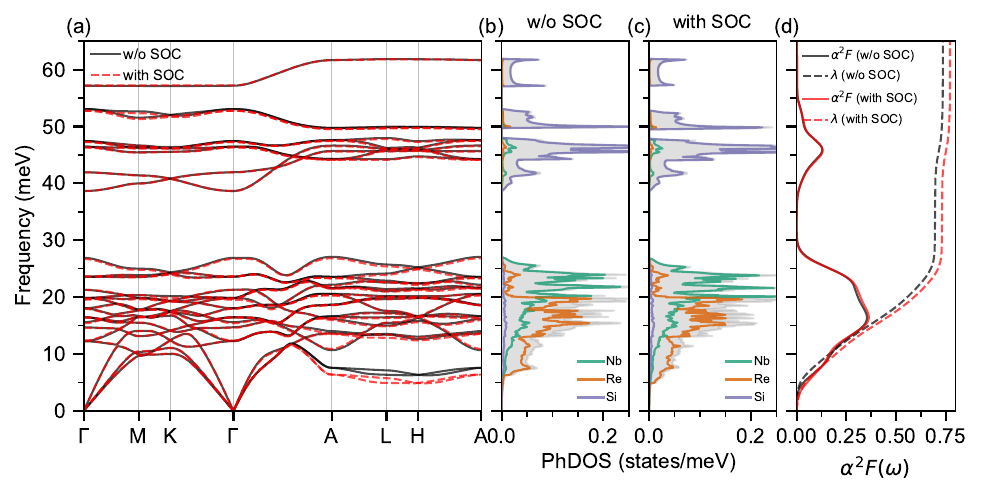}
    \caption{Effect of SOC on the lattice dynamics. (a) Phonon dispersion without (black) and with (red dashed) SOC. (b), (c) Total and atom-projected phonon DOS (Nb teal, Re orange, Si purple) without and with SOC. (d) $\a^2F(\o)$ (solid) and $\l(\o)$ (dashed) for the two cases, on the $4\times4\times8$ DFPT $\bq$-mesh.}
    \label{fig:phonon_soc}
\end{figure}

Figure~\ref{fig:phonon_soc} compares the phonons and $\a^2F(\o)$ without and with spin-orbit coupling (SOC). SOC slightly softens the lowest acoustic branches along A--L--H--A and raises $\l$ from $0.72$ to $0.75$ on the $4\times4\times8$ DFPT $\bq$-mesh (unconverged values; only their $4\%$ ratio is meaningful). Applied to the converged coupling, this raises $\tc$ by less than $0.3$~K, so the main text reports results without SOC.

\section{\label{sec:anh_T}Convergence and temperature dependence of the anharmonic phonons}

\begin{figure}[!hbt]
    \centering
    \includegraphics[width=0.7\linewidth]{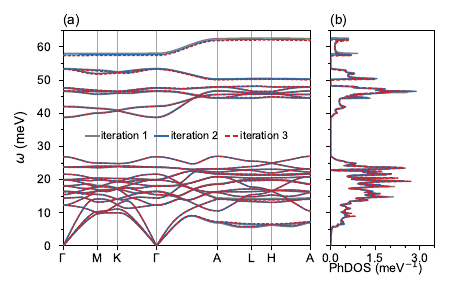}
    \caption{Convergence of the A-SDM iterative cycle at $T=0$~K. (a) Phonon dispersion and (b) phonon DOS.}
    \label{fig:anh_conv}
\end{figure}

Figure~\ref{fig:anh_conv} shows the convergence of the A-SDM cycle: between iterations 2 and 3 the frequencies change by less than $0.3$~meV at all temperatures, and the third iteration is used throughout.


\begin{figure}[!hbt]
    \centering
    \includegraphics[width=0.75\linewidth]{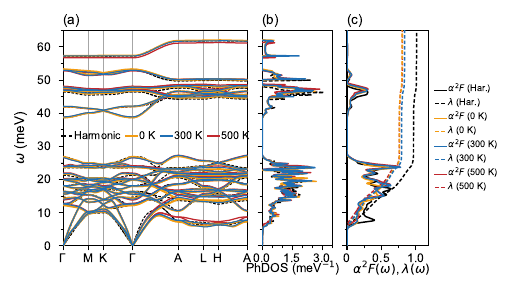}
    \caption{A-SDM phonons at 0, 300, and 500~K (colored) and harmonic phonons (black dashed): (a) dispersion, (b) phonon DOS, (c) $\a^2F(\o)$ (solid) and cumulative $\l(\o)$ (dashed) on the $40\times40\times80$ $\bk$ / $20\times20\times40$ $\bq$ grids.}
    \label{fig:anh_T}
\end{figure}

Figure~\ref{fig:anh_T} compares the A-SDM phonons and $\a^2F(\o)$ at $T=0$, $300$, and $500$~K with the harmonic result. The corresponding couplings are $\l=0.80$, $0.84$, and $0.81$, against the harmonic $\l=1.01$, so the renormalization is dominated by zero-point motion.

\section{\label{sec:re3_anh}Microscopic origin of the Re$_3$ anharmonicity}

Table~\ref{tab:re3} gives the harmonic and anharmonic frequencies and couplings $\l_{\bq\nu}$ of the Re$_3$ branches of Fig.~4 of the main text at A, L, and H. The breathing mode shows the largest anharmonic hardening and loss of coupling. The out-of-plane mode displays an intermediate, wave-vector-dependent response, whereas the weakly coupled rotation shows essentially no response.

\begin{table}[!hbt]
\centering
\caption{Frequencies and mode-resolved couplings $\l_{\bq\nu}$ of the Re$_3$ modes at A$=(0,0,\tfrac12)$, L$=(\tfrac12,0,\tfrac12)$, and H$=(\tfrac13,\tfrac13,\tfrac12)$ with harmonic (Har) and anharmonic (Anh) force constants. The out-of-plane mode is doubly degenerate at A and H.}
\label{tab:re3}
\setlength{\tabcolsep}{4pt}
\renewcommand{\arraystretch}{1.15}
\begin{tabular}{l c c c c c}
\hline\hline
 &  & \multicolumn{2}{c}{$\o_{\bq\nu}$ (meV)} & \multicolumn{2}{c}{$\l_{\bq\nu}$} \\
\cline{3-4}\cline{5-6}
Character & $\bq$ & Har & Anh & Har & Anh \\
\hline
Re$_3$ out-of-plane & A & 7.5  & 8.7  & 0.223 & 0.200 \\
                    & L & 6.4  & 6.6  & 0.138 & 0.161 \\
                    & H & 6.2  & 7.3  & 0.098 & 0.085 \\
Re$_3$ breathing    & A & 13.7 & 15.4 & 0.323 & 0.253 \\
                    & L & 13.4 & 15.0 & 0.292 & 0.244 \\
                    & H & 12.7 & 14.8 & 0.326 & 0.223 \\
Re$_3$ rotation     & A & 16.1 & 16.8 & 0.071 & 0.070 \\
                    & L & 15.0 & 15.0 & 0.027 & 0.027 \\
                    & H & 15.8 & 15.4 & 0.028 & 0.033 \\
\hline\hline
\end{tabular}
\end{table}

Table~\ref{tab:ifc} summarizes the near-neighbor harmonic IFCs in NbReSi. Metal--Si bonds are stiff ($\lVert\bm\Phi\rVert>1$~eV/\AA$^2$), whereas the in-plane Re--Re trimer and kagome interactions are weak, with $\lVert\bm\Phi\rVert=0.50$ and $0.23$~eV/\AA$^2$, and negligible longitudinal components. Removing both in-plane Re--Re IFCs reduces the Re$_3$ breathing-branch dispersion only from $0.92$ to $0.58$~meV, while removing the Re--Re coupling along $c$ changes the $\Gamma$--A dispersion by $<0.5$~meV. Thus, the breathing-mode restoring force arises primarily from Re--Si and Re--Nb interactions rather than direct Re--Re bonding, consistent with its softness and strong anharmonic renormalization.

\begin{table}[!hbt]
\centering
\caption{Near-neighbor harmonic IFCs in NbReSi. Here, $d$ is the bond length and $n$ the coordination multiplicity. The longitudinal IFC,
$\Phi_{\rm L}=\hat{\mathbf u}\cdot\bm\Phi_{ij}\cdot\hat{\mathbf u}$,
measures the force-constant component along the bond direction $\hat{\mathbf u}$, while $\lVert\bm\Phi\rVert$ is the Frobenius norm of the full $3\times3$ IFC block. In-plane metal--metal contacts are italicized.}
\label{tab:ifc}
\setlength{\tabcolsep}{4pt}
\renewcommand{\arraystretch}{1.15}
\begin{tabular}{l c c c c}
\hline\hline
Pair & $d$ (\AA) & $n$ & $|\Phi_{\rm L}|$ (eV/\AA$^2$) & $\lVert\bm\Phi\rVert$ (eV/\AA$^2$) \\
\hline
Re--Si$(1a)$          & 2.43 & 2 & 2.50 & 2.54 \\
Re--Si$(2d)$          & 2.58 & 2 & 2.36 & 2.39 \\
Nb--Si$(2d)$          & 2.67 & 4 & 1.30 & 1.32 \\
Nb--Si$(1a)$          & 2.76 & 1 & 1.12 & 1.15 \\
Nb--Re                & 2.86 & 2 & 0.75 & 0.97 \\
Nb--Re                & 2.94 & 4 & 0.25 & 0.31 \\
Nb--Nb $\parallel c$  & 3.32 & 2 & 1.76 & 1.82 \\
Re--Re $\parallel c$  & 3.32 & 2 & 0.41 & 0.57 \\
\textit{Re--Re (trimer)} & 3.07 & 2 & 0.08 & 0.50 \\
\textit{Nb--Nb (kagome)} & 3.62 & 4 & 0.01 & 0.11 \\
\textit{Re--Re (kagome)} & 4.48 & 4 & 0.04 & 0.23 \\
\hline\hline
\end{tabular}
\end{table}

\section{\label{sec:vertex_origin}Four-state phase space and vertex spectral function}

\begin{figure}[!hbt]
    \centering
    \includegraphics[width=0.75\linewidth]{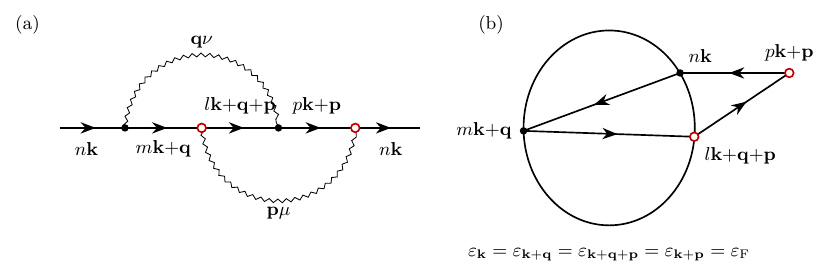}
    \caption{(a) Lowest-order vertex correction to the electron self-energy: two phonons $\bq\nu$ and $\bp\mu$ connect four electronic states; the intermediate states marked by open circles are off-shell in a generic metal~\cite{Mishra2025}. (b) The corresponding scattering process on the Fermi surface. The vertex spectral function $\a^2F^{\rm V}(\o,\op)$ averages over configurations in which all four states lie at $\ef$; their total weight defines $C(0)$.}
    \label{fig:vertexdiag}
\end{figure}

The lowest-order vertex diagram [Fig.~\ref{fig:vertexdiag}(a)] connects four electronic states, $\bk$, $\bk+\bq$, $\bk+\bp$, and $\bk+\bq+\bp$, all of which must lie at the Fermi level for the process to contribute to pairing [Fig.~\ref{fig:vertexdiag}(b)]. In the isotropic theory, the vertex spectral function $\a^2F^{\rm V}(\o,\op)$ averages over these four-state combinations, normalized by their total phase-space weight~\cite{Mishra2025},

\begin{equation}
C(0)=\sum_{nmlp}\sum_{\bk\bq\bp}\d(\ve_{n\bk}-\ef)\,\d(\ve_{m\bk+\bq}-\ef)\,\d(\ve_{l\bk+\bq+\bp}-\ef)\,\d(\ve_{p\bk+\bp}-\ef).
\end{equation}

Defining the Fermi-surface weight $f(\bk)=\sum_n\d(\ve_{n\bk}-\ef)$ for the Fermi-surface weight at $\bk$ and the nesting function $J(\bq)=N_k^{-1}\sum_{\bk}f(\bk)f(\bk+\bq)$ for the nesting function, the triple sum expression reduces to
\begin{equation}
C(0)=\frac{1}{N_k}\sum_{\bq}J(\bq)^2,
\end{equation}

We evaluate $C(0)$ by fast Fourier transform on the DFT $24\times24\times48$ mesh, using $30$~meV Gaussian broadening. The normalized quantity $C(0)/\NF^4$ measures the enhancement of the four-state phase space relative to independently distributed Fermi-level states, for which it is approximately unity. The full Fermi surface of NbReSi gives $C(0)/\NF^4=1.0$, indicating no global nesting enhancement. In contrast, restricting the analysis to states within $0.3$~\AA$^{-1}$ of the high-symmetry H-point, which contribute $18\%$ of $N(\ef)$, increases the ratio to $4.4$. Thus, the enhanced vertex phase space is localized to the nearly dispersionless Re-$d$ band near H, where the four Fermi-level constraints can be satisfied over a broad range of $\bq$ and $\bp$. The appreciable vertex correction in NbReSi therefore originates from this local flat-band phase space, together with the large $\NF$, rather than from global Fermi-surface nesting.

\begin{figure}[!hbt]
    \centering
    \includegraphics[width=0.6\linewidth]{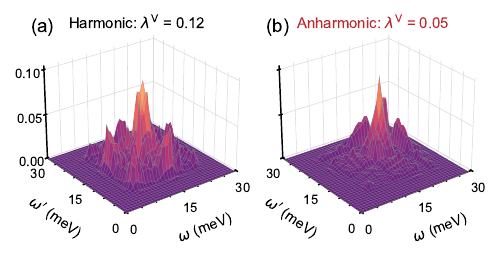}
    \caption{Vertex spectral function $\a^2F^{\rm V}(\o,\op)$ of NbReSi for (a) harmonic and (b) anharmonic phonons, and the integrated vertex coupling $\lV$.}
    \label{fig:a2fv}
\end{figure}

Figure~\ref{fig:a2fv} shows the resulting vertex spectral function $\a^2F^{\rm V}(\o,\op)$ for harmonic and anharmonic phonons. Its weight lies below $30$~meV, within the Nb/Re phonon manifold, and the anharmonic hardening of the lowest Re$_3$ branches reduces the integrated vertex coupling from $\lV=0.12$ to $0.05$; the $1/(\o\op)$ weighting in $\lV$ makes it more sensitive than $\l$ to that hardening.

\section{\label{sec:mustar}Dependence on the Coulomb pseudopotential}

\begin{figure}[!hbt]
    \centering
    \includegraphics[width=0.8\textwidth]{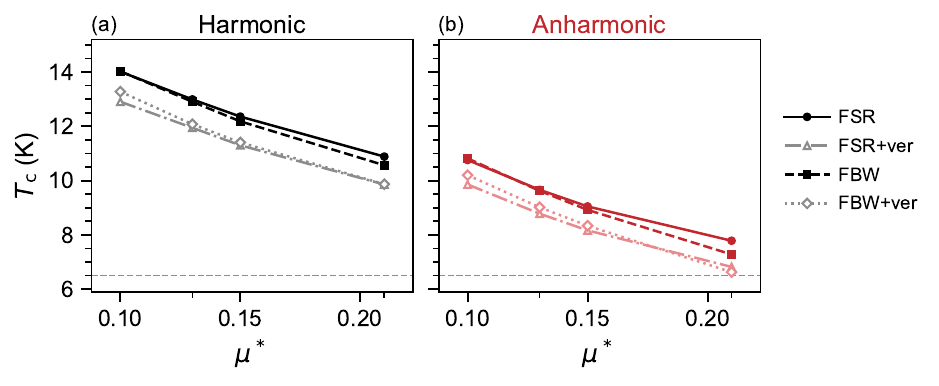}
    \caption{Isotropic Eliashberg $\tc$ of NbReSi versus $\mu^*$ for (a) harmonic and (b) anharmonic phonons at the FSR, FBW, FSR+ver, and FBW+ver levels. Eliashberg equations were solved at $\mu^*=0.10$, $0.13$, $0.15$ and $0.21$. }
    \label{fig:tc_mustar}
\end{figure}

Figure~\ref{fig:tc_mustar} and Table~\ref{tab:mustar} give the $\mu^*$ dependence. Over $\mu^*=0.10$--$0.21$, the hierarchy of corrections is unchanged, and FSR and FBW differ by less than $0.5$~K; at the conventional $\mu^*=0.15$, the anharmonic vertex-corrected $\tc=8.2$~K still exceeds the experimental $\tc\simeq6.5$~K of hexagonal NbReSi~\cite{Su2021,Shang2022}, which motivates $\mu^*=0.21$, the value required for elemental Nb~\cite{Savrasov1996}.

\begin{table}[!hbt]
\centering
\caption{Superconducting properties of NbReSi versus $\mu^*$ for harmonic and anharmonic phonons. $\tc^{\rm MAD}$ denotes the Allen--Dynes modified McMillan estimate~\cite{Allen1975}, while FSR and FBW denote isotropic Eliashberg results without and with (+ver) vertex corrections. $\Delta(0)$ and $2\Delta(0)/k_{\rm B}\tc$ are given for the FSR solution. Experimental values are for hexagonal polycrystals, from Refs.~\cite{Su2021,Shang2022}.}
\label{tab:mustar}
\setlength{\tabcolsep}{5pt}
\renewcommand{\arraystretch}{1.2}
\begin{tabular}{l c c c c c c c c}
\hline\hline
Phonons & $\mu^*$ & $\tc^{\rm MAD}$ & $\tc^{\rm FSR}$ & $\tc^{\rm FBW}$ & $\tc^{{\rm FSR}+{\rm ver}}$ & $\tc^{{\rm FBW}+{\rm ver}}$ & $\Delta(0)$ & $2\Delta(0)/k_{\rm B}\tc$ \\
\hline
Harmonic & 0.10 & 12.8 & 14.0 & 14.0 & 12.9 & 13.3 & 2.40 & 3.98 \\
Harmonic & 0.13 & 10.9 & 13.0 & 12.9 & 11.9 & 12.1 & 2.20 & 3.92 \\
Harmonic & 0.15 & 9.7 & 12.4 & 12.2 & 11.3 & 11.4 & 2.08 & 3.90 \\
Harmonic & 0.21 & 6.5 & 10.9 & 10.6 & 9.9 & 9.9 & 1.80 & 3.84 \\
\hline
Anharmonic & 0.10 & 9.5 & 10.8 & 10.8 & 9.9 & 10.2 & 1.75 & 3.78 \\
Anharmonic & 0.13 & 7.6 & 9.7 & 9.6 & 8.8 & 9.0 & 1.56 & 3.75 \\
Anharmonic & 0.15 & 6.5 & 9.0 & 8.9 & 8.2 & 8.3 & 1.45 & 3.73 \\
Anharmonic & 0.21 & 3.6 & 7.8 & 7.3 & 6.8 & 6.6 & 1.20 & 3.58 \\
\hline
Experiment & -- & -- & \multicolumn{4}{c}{$6.5$} & $0.98$--$1.04$ & $3.50$--$3.90$ \\
\hline\hline
\end{tabular}
\end{table}

\newpage

\bibliography{pap}